\documentstyle[12pt,epsfig]{article}
\def\beq{\begin{equation}}
\def\eeq{\end{equation}}
\def\bea{\begin{eqnarray}}
\def\eea{\end{eqnarray}}
\def\nnb{\nonumber}

\newcommand{\gsim}{\lower.7ex\hbox{$\;\stackrel{\textstyle>}{\sim}\;$}}
\newcommand{\lsim}{\lower.7ex\hbox{$\;\stackrel{\textstyle<}{\sim}\;$}}

\begin{document}
%\twocolumn[\hsize\textwidth\columnwidth\hsize\csname@twocolumnfalse\endcsname
\begin{flushright}
\baselineskip=12pt
UPR-1005-T, IC/2002/68\\
%hep-ph/
\end{flushright}
\vskip 0.2cm

\begin{center}
{\LARGE\bf

Low Energy Gauge Unification Theory}

\vspace{1cm}
%\author{	
{\Large
Tianjun Li$^a$ and  Wei Liao$^b$}\\
\vspace{0.4cm}
%\address{
        $^a$ Department of Physics and Astronomy,\\ 
University of Pennsylvania, Philadelphia, PA 19104-6396, USA \\
        $^b$ ICTP, Strada Costiera 11, 34014 Trieste, Italy
%}
\end{center}
%\maketitle
\begin{abstract}
Because of the problems arising from the fermion unification in
the traditional Grand Unified Theory and the mass hierarchy
between the 4-dimensional Planck scale and weak scale, 
we suggest the low energy gauge unification theory with low
high-dimensional Planck scale. We discuss
the non-supersymmetric $SU(5)$ model on $M^4\times S^1/Z_2 \times S^1/Z_2$
and the supersymmetric $SU(5)$ model on $M^4\times S^1/(Z_2\times Z_2')
\times S^1/(Z_2\times Z_2')$. The $SU(5)$ gauge symmetry is broken
by the orbifold projection for the zero modes, and the gauge unification
is accelerated due to the $SU(5)$ asymmetric light KK states.
In our models, we forbid the proton decay, still keep the
charge quantization, and automatically solve the fermion mass
problem. We also comment on the anomaly cancellation and
 other possible scenarios for
low energy gauge unification.
\vskip 0.2cm
\end{abstract}
%]

\section{Introduction} \label{sec1}
The idea of a Grand Unified Theory (GUT) since its birth~\cite{ggqw} has
been so attractive that by now GUTs are widely considered as the very good extensions
of the Standard Model (SM), especially when we consider the supersymmetry.
 One impressive
success of this idea is that including the radiative corrections,
GUTs give  
the correct weak mixing angle, which is observed in the electroweak (EW) scale
experiments. The confusing problem of the $U(1)$ charge quantization
is also resolved through embedding the SM hypercharge symmetry into the
non-abelian gauge symmetry. Another nice point of GUT is the so-called
fermion unification, {\it i. e.}, the SM fermions of each generation elegantly
fit the $\bar{5}+10$ representation of the $SU(5)$ group, or the $16$ representation of
the $SO(10)$ group if we included the right handed neutrinos.

Although these nice points seem quite encouraging to the theoretical physicists
 who pursue the idea
of GUT, fermion unification always gives us problems. In spite of the
supporting evidence of the observed $m_\tau \approx m_b$, GUT
gives the wrong fermion mass relation $m_e/m_{\mu} = m_d/m_s$
at 1 GeV scale,  which is not supported
by experiments. This fermion mass problem may be solved by introducing
the Higgs whose dimension is larger than that of the adjoint representation.
 And the fermion unification naturally leads to the notable proton
decay problem. The broken gauge bosons (X and Y in the $SU(5)$ gauge
group) can lead to the proton decay through the dimension six operator,
and the present experiments pushed them to be at least as massive as $5\times
10^{15}$ GeV~\cite{sk}. For the supersymmetric GUTs,
this bound may not be harmful because the unfication scale is about $2.4\times 10^{16}$
GeV, however, there
exists dimension five proton decay operator by
higgsinos exchange. This raises the famous Higgs
doublet-triplet spliting problem.
One may soon realise that it's not easy to solve this
doublet-triplet splitting problem without including the complicated
interactions and fine-tuning.
In short, we do have enough problems from the fermion unification.

Recently, grand unified theories have been revisited in the framework
with extra dimensions~\cite{kawa,alot,LTJ1,LTJ2,hn}.
Using the orbifold projection, the $5$ (or $\bar{5}$) Higgs fields living 
in the high dimensional bulk can be naturally splitted into massive
triplets and light doublets. Furthermore, it was shown that all the dimension 
four and five proton decay operators may be completely avoided because
of the intrinsic global symmetry~\cite{hn}. But in this kind of models, the
$10^{15}$ GeV compactification scale is naturally involved and
the dimension six proton decay operator may give disastrous consequence, so,
one has to sacrifice the unification of the first
generation fermions. Because of the SUSY flavor problem, the unification
of the second generation fermion is also sacrificed.
Of course, the wrong prediction of $m_e/m_{\mu} = m_d/m_s$ is avoided
simultaneously~\cite{hn}.

Because of the difficulties in searching for a natural grand unified theory,
we might conclude that the
fermion unification in the usual form familiar to us always 
leads to the serious problems. Therefore, we may conjecture that
the realistic fundamental theory, which describes the nature, might be the
theory with the gauge unification and without the fermion unification.
In addition, in the gauge unification theory, we might suppress the proton
decay~\cite{LTJ1}. 
Then, the scale of gauge unification can be as low as hundreds
of TeV if it is possible.
 This kind of scenario is very interesting
if the string scale or high-dimensional Planck scale is
at hundreds of TeV scale range, and then, there does not
exist the mass hierarchy between the high-dimensional Planck
scale and unification scale, although there exists
the mass hierarchy between the 4-dimensional Planck scale and
weak scale. As a comment, let us briefly 
discuss how to low the high-dimensional Planck scale.
 Suppose we have 10-dimensional space-time, the gauge
fields live on a 5-brane, the 4-dimensional Planck scale $M_{Pl}$ is
related to the 10-dimensional Planck scale $M_*$ and the
physical size of the extra dimensions as follows
\bea
M^2_{P} \sim M_*^8 (R_1 R_2) {\tilde R}^4,
\eea
where $R_1$ and $R_2$ are the two radii of the extra dimension where
gauge fields can propagate, ${\tilde R}$ is the common radius of the
extra four space-like extra dimensions. Assuming $M_* \sim 100$ TeV and
$R_1,R_2 \sim 10$ TeV, we get $1/{\tilde R} \sim 0.1/{\textrm GeV}$.

Moreover,
it is widely believed that even in 4-dimensional theory, the CKM mixing is a direct
consequence of the string scale physics. In the approach without fermion
unification, we therefore have to ascribe not only the flavor mixing but
also all the Yukawa couplings directly to the string scale physics.
This idea is further pushed by the observation that involving a
TeV scale extra dimension, the accumulating KK excitations may accelerate
the gauge coupling running, and low the unification scale,
$M_U$, even to TeV scale~\cite{ddg}. 

To get an idea on how the KK excitations work for our purpose, let
us remind the readers in the following way. In 4-dimensional theory, the runnings
of the gauge couplings are
\bea
\alpha_i^{-1}(\mu) = \alpha_i^{-1}(M) + \frac{b_i}{4 \pi} ln\frac{\mu^2}{M^2}, 
\label{run1}
\eea
where $\alpha_i=g_i^2/4\pi$, $i=1,2,3$ refers to $SU(3)_C\times SU(2)_L
\times U(1)_Y$ gauge groups, $b_i$ is the corresponding beta function.
$M$ should be the energy scale not
much higher than $m_Z$, for example $M_{SUSY}$ in MSSM which is
of TeV scale. Because of the big gap from the TeV scale to the 
4-dimensional unification scale
 ($10^{16}$ GeV), changing $M$ around TeV scale is just a tiny change on the
boundary conditions and does not affect the precision of gauge
coupling unification. Beginning with the initial value of
$\alpha_i$ at the EW scale, whether the gauge couplings may unify to
be $SU(5)$ symmetric depends on the differences among the beta functions
$(b_2-b_1, b_3-b_2)$. In the SM or MSSM, the gauge and Higgs sectors
contribute to the differences of the
beta functions while the matter contents do not, because
the matter contents are completely $SU(5)$ symmetric. If $N$ copies of gauge
and Higgs sectors involved not far from the EW scale, we then have
\bea
\alpha_i^{-1}(\mu) -\alpha_j^{-1}(\mu) = \alpha_i^{-1}(M) -\alpha_j^{-1}(M)
+N \frac{b_i-b_j}{4 \pi} ln\frac{\mu^2}{M^2}.
\label{run2}
\eea
The unification scale can be lowed in this approach, for instance, 
 $M_U$ can be around tens of TeV if $N=13$~\cite{ach}. It's
clear that the precision of gauge coupling unification is of the
same level as the original one given by Eq. (\ref{run1}). In the
framework of large extra dimension, for example with one extra
space-like dimension, we can effectively take the point of view of an 
4-dimensional
 truncated Kaluza-Klein theory and the story is similar but a little bit
different. Because of
the particle content is accumulated as the energy increased, we have
\bea
\alpha_i^{-1}(\mu) -\alpha_j^{-1}(\mu) = \alpha_i^{-1}(M_c) -\alpha_j^{-1}(M_c)
+\frac{b^0_i-b^0_j}{4 \pi} ln\frac{\mu^2}{M_c^2}
+\sum_n \frac{b^n_i-b^n_j}{4 \pi} ln\frac{\mu^2}{M^2_n}, 
\label{run23}
\eea
where $M_c$ is the compactification scale, $M_n$ is the mass of the KK
states, $n > 0$ refers the massive KK levels, $b^0_i$ refers the beta function
given by the zero mode. In the simple case $b^n_i-b^n_j=b_i-b_j$, we get
\bea
\alpha_i^{-1}(\mu) -\alpha_j^{-1}(\mu) = \alpha_i^{-1}(M_c) -\alpha_j^{-1}(M_c)
+\frac{b^0_i-b^0_j}{4 \pi} ln\frac{\mu^2}{M_c^2}
+\frac{b_i-b_j}{4 \pi} ln\prod_n\frac{\mu^2}{M^2_n}, 
\label{run3}
\eea
This formula then leads to another unification scale $M_U$ which is just one
order of magnitude larger than the compactification scale. 
The gauge coupling unification can be achieved at hundreds of TeV
if a 10 TeV compactification scale is involved.
If $b^0_i-b^0_j = b_i-b_j$, 
similar to the above example, we have the gauge coupling
unification.
If $b^0_i-b^0_j \not =  b_i-b_j$, one may also realise that
$b^0_i-b^0_j$ can not affect too much on the mismatch
of the gauge couplings at the scale $M_U$, because the contribution
to the gauge coupling running given by the massive KK states is
one order of magnitude larger than the zero mode contribution.
For example, the SM beta functions are $(b_1,b_2,b_3)=(-41/10,19/6,7)$, and
the differences among them are $(b_2-b_1,b_3-b_2)=(109/15,23/6)$. As we know, it
can not unify the gauge couplings in the 4-dimension and it can not work
in the high dimensional models either unless we introduce
the extra matter contents, which will be discussed in
the section 2. In addition, the MSSM beta functions are
$(-33/5,-1,3)$,  and they give the differences $(b_2-b_1,b_3-b_2)=(28/5,4)$.
In 5-dimensional theory, this choice can 
work only if the approriate particle contents were involved in the high dimensional
bulk. In section 3, we will give two scenarios
to recover this MSSM differences among the beta functions
 in the high dimensional supersymmetric model.

In this letter, we will focus on the 6-dimensional
space-time or the 5-brane, and try to build the
$SU(5)$ models with low energy gauge unification and without fermion unification.
We discuss the non-supersymmetric $SU(5)$ model on 
the space-time $M^4\times S^1/Z_2 \times S^1/Z_2$
where there are three pairs of Higgs doublets in the bulk;
the supersymmetric $SU(5)$ model on 
the space-time $M^4\times S^1/(Z_2\times Z_2')
\times S^1/(Z_2\times Z_2')$ where there are one
pair of Higgs doublets and one pair of hypercharge one singlets
on the boundary 4-brane at $z=\pi R_2/2$, or one Higgs doublet ($H_d$)
on the boundary 4-brane at $z=\pi R_2/2$. In addition, 
we forbid the proton decay by putting the matter fields on
the suitable 3-branes at the fixed points, and we
still have the charge quantization and automatically avoid the
fermion mass problem in our models.
The gauge unfication scale is about hundreds of TeVs,
and we would like to point out that
in order to avoid the uncertainties from the loop corrections
to the masses of the KK states, the supersymmetry breaking soft
parameters and Higgs mechanism, we consider the scenario with
$R_1 \sim 10 R_2$, and then the gauge unification is accelerated by
the leading order power one running. We would like to emphasize
that for the scenarios with $R_1 \sim R_2$, we still have
 low energy gauge unification, but the gauge unification is
accelerated by the subleading order contributions
which might be affected by the uncertainties. These uncertainties
might be small and definitely deserve further study.

\section{Non-Supersymmetric Model} \label{sec2}

In this section, we would like to discuss the non-supersymmetric
$SU(5)$ model on the space-time $M^4 \times S^1/Z_2 \times S^1/Z_2$.
 The model is given as Model II in Ref.~\cite{LTJ1}, where the
gauge unification was not discussed. In this model, 
the gauge symmetry $SU(5)$ is broken down to the Standard Model
gauge symmetry by orbifold projection, and 
we only have the zero modes and KK modes of the
Standard Model gauge fields and two Higgs doublets on the observable 3-brane
 at the fixed point.
 So, we can have the low energy unification, and solve 
the doublet-triplet splitting problem,
 the gauge hierarchy problem, and the proton decay problem.
Let us review the set-up.

We consider 
the 6-dimensional space-time which can be factorized into the product of the 
ordinary 4-dimensional space-time $M^4$ and the orbifold 
$S^1/Z_2 \times S^1/Z_2$. The corresponding
coordinates are $x^{\mu}$, ($\mu = 0, 1, 2, 3$),
$y\equiv x^5$ and $z \equiv x^6$.
The radii for the circles along $y$ direction and $z$ direction are
$R_1$ and $R_2$, respectively. The orbifold 
$S^1/Z_2 \times S^1/Z_2$ is obtained by $S^1\times S^1$ moduloing the equivalent
classes: $y \sim -y$ and $ z \sim -z$. 
So, we have four fixed points: $(y=0,z=0)$, $(y=0,z=\pi R_2)$, $(y=\pi R_1, z=0)$
and $(y=\pi R_1, z= \pi R_2)$ that are fixed under two actions, and
four fixed lines: $y=0$, $y=\pi R_1$, $z=0$ and $z=\pi R_2$
that are fixed under one action.
 We associate two 
parity transformations to the fields living in the bulk. For an example,
for a field $\phi(x^\mu,y,z)$ in the fundamental representation we have
\bea
\phi(x^\mu,y,z) \to \phi(x^\mu,-y,z)=\eta_{\phi}^y P \phi(x^\mu,y,z), ~
\phi(x^\mu,y,z) \to \phi(x^\mu,y,-z)=\eta_{\phi}^z P^\prime \phi(x^\mu,y,z)~,
\label{parity1}
\eea
where $\eta_{\phi}^y$ and $\eta_{\phi}^z$ are $\pm1$.

And we assume the trivial periodic boundary condition for all
the fields involved in the model:
\bea
\phi(x^\mu,y+2 \pi R_1,z)=\phi(x^\mu,y,z+2 \pi R_2)=\phi(x^\mu,y,z)~.
\eea

Denoting the field with parities $(\pm,\pm)$ under $(P,P^\prime)$
by $\phi_{\pm \pm}$, we obtain the
corresponding Fourier expansions~\cite{LTJ1}: $\cos{n y \over R_1} \cos{m z \over R_2}$
for $\phi_{++}$, $\cos{n y \over R_1} \sin{m z \over R_2}$ for $\phi_{+-}$,
$\sin{n y \over R_1} \cos{m z \over R_2}$ for $\phi_{-+}$ and
$\sin{n y \over R_1} \sin{m z \over R_2}$ for $\phi_{--}$, where
$n$ and $m$ are non-negative integers. As expected
$\phi_{-+}$ and $\phi_{--}$ vanish at $y=0$ and $\pi R_1$,
and $\phi_{+-}$ and $\phi_{--}$ vanish at $z=0$ and $\pi R_2$.
The masses of these KK modes are given by $\sqrt{n^2/R^2_1+m^2/R_2^2}$,
$\sqrt{n^2/R^2_1+(m+1)^2/R^2_2}$, $\sqrt{(n+1)^2/R^2_1+m^2/R^2_2}$
and $\sqrt{(n+1)^2/R^2_1+(m+1)^2/R^2_2}$ ($n,m \ge 0$) for fields of parities
$(+,+)$, $(+,-)$, $(-,+)$ and $(-,-)$, respectively.

We choose the following parity assignments for $P$ and $P^\prime$:
\bea
P={\rm diag} \{ -1, -1, -1, 1,1\}, ~
P^\prime={\rm diag} \{-1,-1,-1,1,1\}.
\label{parity2}
\eea
Upon the parity transformation, the gauge generators of $SU(5)$
group are seperated into two classes: the generators of the SM
gauge group $T^a$, and the other broken gauge generators $T^{\hat a}$,
and $T^a$ and $T^{\hat a}$ satisfy the following equations
\bea
P T^a P^{-1} = P^\prime T^a P^{\prime -1}= T^a, ~
P T^{\hat a} P^{-1} = P^\prime T^{\hat a} P^{\prime -1}= - T^{\hat a}.
\eea

\renewcommand{\arraystretch}{1.4}
\begin{table}[tt]
\caption{Parity assignment and masses ($n \ge 0, m \ge 0$ ) 
of the fields in the SU(5) gauge
 and Higgs multiplets for the non-supersymmetric model.
\label{tab:chiralI}}
\vspace{0.4cm}
\begin{center}
\begin{tabular}{|c|c|c|}
\hline    
$(P,P')$ & field & mass\\ 
\hline
$(+,+)$ &  $A^a_{\mu}$, $H^D_u$, $H^D_d$ & $\sqrt {n^2/R_1^2+m^2/R_2^2}$\\
\hline
$(+,-)$ &  $A^{\hat{a}}_{5}$, $A^{a}_6$, $H_i^T$ 
 & $\sqrt {n^2/R_1^2+(m+1)^2/R_2^2}$  \\
\hline
$(-,+)$ &  $A^{a}_5$, $A^{\hat{a}}_6$, $H_i^D$ 
& $\sqrt {(n+1)^2/R_1^2+m^2/R_2^2} $\\
\hline
$(-,-)$ &  $A^{{\hat a}}_{\mu}$, $H^T_u$, $H^T_d$ & 
$\sqrt {(n+1)^2/R_1^2+(m+1)^2/R_2^2} $ \\
\hline
\end{tabular}
\end{center}
\end{table}

\renewcommand{\arraystretch}{1.4}
\begin{table}[t]
\caption{The gauge fields, Higgs fields and gauge group
on the 3-branes that are located
at the fixed points $(y=0, z=0),$ $(y=0, z=\pi R_2)$,
$(y=\pi R_1, z=0)$ and $(y=\pi R_1, z=\pi R_2)$, and
on the 4-branes that are located at $y=0$, $z=0$, $y=\pi R_1$
and $z=\pi R_2$.}
\label{tab1}
\vspace{0.4cm}
\begin{center}
\begin{tabular}{|c|c|c|}
\hline 
Brane position & field & gauge group\\
\hline
$(y=0, z=0)$ & $A_{\mu}^a$, $H_u^D$, $H_d^D$ & ~~ $SU(3)\times SU(2)\times
U(1)$ ~~ \\
\hline
$(y=0, z=\pi R_2)$ & $A_{\mu}^a$, $H_u^D$, $H_d^D$ & $SU(3)\times
SU(2)\times U(1)$ \\
\hline
$(y=\pi R_1, z=0)$ & $A_{\mu}^a$, $H_u^D$, $H_d^D$ & $SU(3)\times
SU(2)\times U(1)$ \\
\hline
~~ $(y=\pi R_1, z=\pi R_2)$ ~~& $A_{\mu}^a$, $H_u^D$, $H_d^D$ &
$SU(3)\times SU(2)\times U(1) $ \\
\hline
$y=0$ or $y=\pi R_1$ &  ~~ $A_{\mu}^a$, $A^{\hat{a}}_5$, $A^{a}_6$, $H_u^D$, $H_d^D$
$H^T_i$ ~~ & $SU(3)\times SU(2)\times U(1)$ \\
\hline
$z=0$  or $z=\pi R_2$ & ~~ $A_{\mu}^a$, $A^{a}_5$, $A^{\hat{a}}_6$, $H_u^D$, $H_d^D$
$H_i^D$ ~~ & $SU(3)\times SU(2)\times U(1)$ \\
\hline
\end{tabular}
\end{center}
\end{table}

Now let us discuss gauge coupling unification.
Notice that even if $R_1 \sim R_2$ involved, it is
still possible to get the gauge coupling unification, because
the orbifold projection makes parts of the fields disappear
in some of the KK states and then the relative running can
accelerated. However in the case of $R_1 \sim R_2$, the dominating
radiative corrections to the gauge couplings are basically
$SU(5)$ symmetric because of the $SU(5)$ bulk theory, and 
the relative running is given by the subleading order contributions
which depend largely on the differences of the beta functions
and detailed masses of different fields. If including 
the loop corrections to the masses of the KK states, the Higgs
mechanism to break EW symmetry (and the supersymmetry breaking soft
parameters in the supersymmetric case), it might be possible that the precision
on the gauge coupling unification would be changed a lot.
Therefore, we consider the scenario with
$R_1 \sim 10 R_2$ and make the relative running to be
at leading order in the whole radiative corrections. 
If the relative running is at the leading order
as the form of the last term in Eq. (\ref{run3}), it's clear
that uncertainties in the masses of the KK states can not
change the precision of the gauge coupling unification although they
may affect the scale a little bit
 at which the gauge coupling unification is reached.

At the energy scale $ 1/R_1 < \mu < 1/R_2$, from the model described 
above we would have a 5-dimensional effective theory with the SM gauge 
group in the 5-dimensional bulk and at each KK level,
there are states coming from $A^a_5$ and $A^{\hat a}_6$ fields.
To achieve gauge coupling unification we add three pairs
of two Higgs ($5$ and ${\bar 5}$) in the 6-dimensional bulk.
\footnote{Putting one pair of Higgs doublets in the bulk is not enough.}
We may use the orbifold projection to eliminate the zero modes of
two pairs of them from
the 4-dimensional effective theory. For one pair
 of them, denoting as $H_u$ and $H_d$,
 the parity transformations are given by Eq. (\ref{parity1}) and 
(\ref{parity2}), and $\eta_{\phi}^y=+1$ and $\eta_{\phi}^z=+1$.
 For the other two pairs, denoting as
$H_i$ where $i=1, 2, 3, 4$, 
we choose $\eta_{\phi}^y=-1$ and $\eta_{\phi}^z=+1$.
The parities and mass spectrum of the gauge fields and Higgs
fields are given  in the Table 1. And in Table 2,  we present
the gauge fields, Higgs fields and gauge group
on the 3-branes that are located
at the fixed points $(y=0, z=0),$ $(y=0, z=\pi R_2)$,
$(y=\pi R_1, z=0)$ and $(y=\pi R_1, z=\pi R_2)$, and
on the 4-branes that are located at $y=0$, $z=0$, $y=\pi R_1$
and $z=\pi R_2$.

From Table 1,  we obtain that the triplet Higgs do not have zero and
light KK modes while the doublet Higgs do have light KK modes.
In short, in the effective
5-dimensional theory, we have the SM gauge fields, adjoint scalar fields
$A^a_5$, lepton-quark scalar $A^{\hat a}_6$ with hypercharge $5/6$ and 
three pairs of two Higgs doublets.
For each KK level below the scale $1/R_2$, $(b_1,b_2,b_3)=
(-{43 \over 30},{11 \over 2}, {21 \over 2})$,
$(b_2-b_1,b_3-b_2)=(104/15,14/3)$
and indeed it can lead to the gauge coupling unification.
We assume the energy scale where the gauge
coupling unified is the cutoff scale, $\Lambda$, of the theory.
For example, assuming $1/R_2=6/R_1$, we find that
the gauge coupling unification
can be achieved at $\Lambda \approx 190$ TeV for $1/R_1=10$ TeV.
 Running above the energy scale $1/R_2$ can also reduce the
discrepancies among the gauge couplings considerately because although
the contributions from those KK states with
masses $\sqrt{n^2/R^2_1+ m^2/R_2^2}$ in which $m > 0$ are 
basically $SU(5)$ symmetric,
there exist the dominant contributions to the
relative gauge couplings running from the KK states with
mass $n/R_1$, {\it i. e.}, $m=0$. The above example with
$\Lambda R_1 \approx 19$ and $\Lambda R_2 \approx 3$ confirms
our expectations, that is the $SU(5)$ asymmetric contributions
are indeed at the leading order.

From Table 2, we know that on the 3-brane at any one of
four fixed points, there exist only the SM gauge fields 
and one pair of Higgs doublets $H_u$ and $H_d$.
 We can then put the SM fermions on one of the 3-branes, for
example, the 3-brane at $(y=0,z=0)$. This theory is completly anomaly-free,
and there are no operators which can lead to the proton decay, as discussed
in the Ref.~\cite{LTJ1}. We can couple the
 Higgs doublets, $H_u$ and $H_d$, to the SM fermions,
like the Model II of the two Higgs doublet Model. Thus, 
the charge quantization is achieved from the consistent condition if 
we included the
Yukawa couplings on the observable 3-brane because the $U(1)$ charge has to be 
balanced in the Yukawa couplings. And together with the four anomaly-free
conditions, we can determine the $U(1)$ charges of the SM fermions. 

In short, the physical
picture in our model is that above the low compactification scale ($1/R_1$), 
$SU(3)_C \times SU(2)_L \times U(1)_Y$ gauge fields and three pairs of Higgs
doublets propagate
in the 5-dimensional space-time,  and the light KK states with masses 
$n/R_1$ accelerate
the gauge coupling unification. Above the high compactification scale ($1/R_2$),
 we should have the $SU(5)$ gauge theory,  
and the gauge coupling unification is achieved at the cutoff scale.
So, at least two extra dimensions in this kind of models should be involved. 

\section{Supersymmetric Model} \label{sec3}
In this section, we would like to discuss the low energy supersymmetric
$SU(5)$ model, which is the supersymmetric extension of the model
in the last subsection.

One of the two possibilities is that we consider the 6-dimensional
$N=1$ supersymmetric theory. In terms of the 4-dimensional language, this theory
has $N=2$ supersymmetry and the gauge superfield is described by a vector superfield
$V$ and a chiral superfield $\Phi$. The scalar component of $\Phi$
can be written as $A_5+ i A_6$. However, it is not hard for one to figure out
that we can not construct the 6-dimensional $N=1$ supersymmetric $SU(5)$ model,
 which is the supersymmetric generalization of the model
in the last subsection, because the discrete symmetry should act on
 $A_5$ and $A_6$ simultaneously and $R_1=R_2$ is required.
Thus, we have to consider the 6-dimensional $N=2$ supersymmetric theory.

The 6-dimensional $N=2$ supersymmetric 
theory is anomaly-free in the bulk. In terms of
 4-dimensional language, the 
6-dimensional $N=2$ supersymmetric theory corresponds to the 4-dimensional
$N=4$ supersymmetric theory. The gauge superfield can be decomposed
to be one vector superfield, $V$, and three chiral superfields, $\Sigma_5$,
$\Sigma_6$ and $\Phi$. 
In the Wess-Zumino gauge and 4-dimensional $N=1$ language, the bulk action
is~\cite{nahgw}
\begin{eqnarray}
  S &=& \int d^6 x \Biggl\{
  {\rm Tr} \Biggl[ \int d^2\theta \left( \frac{1}{4 k g^2}
  {\cal W}^\alpha {\cal W}_\alpha + \frac{1}{k g^2}
  \left( \Phi \partial_5 \Sigma_6 - \Phi \partial_6 \Sigma_5
  - \frac{1}{\sqrt{2}} \Phi
  [\Sigma_5, \Sigma_6] \right) \right) + {\rm h.c.} \Biggr]
\nonumber\\
&&  + \int d^4\theta \frac{1}{k g^2} {\rm Tr} \Biggl[
  \sum_{i=5}^6 \left((\sqrt{2} \partial_i + \Sigma_i^\dagger) e^{-V}
  (-\sqrt{2} \partial_i + \Sigma_i) e^{V} +
   \partial_i e^{-V} \partial_i e^{V}\right)
%\nonumber\\
%  \qquad \qquad \qquad
  + \Phi^\dagger e^{-V} \Phi e^{V}  \Biggr] \Biggr\}.~\,
\label{6daction}
\end{eqnarray}

\renewcommand{\arraystretch}{1.4}
\begin{table}[t]
\caption{Parity assignment and masses ($n\ge 0, m \ge 0$) for 
the vector multiplet in the supersymmetric $SU(5)$ model on 
$M^4 \times S^1/(Z_2\times Z^\prime_z)\times S^1/(Z_2\times Z^\prime_2)$.
And we include the Higgs superfields $(H, H^c)$ on the fixed 4-brane 
at $z=\pi R_2/2$.
\label{tab:SUV1}}
\vspace{0.4cm}
\begin{center}
\begin{tabular}{|c|c|c|}
\hline        
$(P^y, P^{y'}, P^z, P^{z'})$ & field & mass\\ 
\hline
$(+, +, +, +)$ &  $V^a_{\mu}$ & $\sqrt {(2n)^2/R_1^2+ (2m)^2/R_2^2}$ \\
\hline
$(+,-, +, -)$ &  $V^{\hat{a}}_{\mu}$ & $\sqrt {(2n+1)^2/R_1^2+(2m+1)^2/R_2^2}$ \\
\hline
$(-, -, +, +)$ &  $\Sigma_5^a$ & $\sqrt {(2n+2)^2/R_1^2+ (2m)^2/R_2^2}$ \\
\hline
$(-, +, +, -)$ &  $\Sigma_5^{\hat{a}}$ & $\sqrt {(2n+1)^2/R_1^2+ (2m+1)^2/R_2^2}$ \\
\hline
$(+, +, -, -)$ &  $\Sigma_6^a$ & $\sqrt {(2n)^2/R_1^2+ (2m+2)^2/R_2^2}$\\
\hline
$(+, -, -,  +)$ &  $\Sigma_6^{\hat{a}}$ & $\sqrt {(2n+1)^2/R_1^2+ (2m+1)^2/R_2^2}$ \\
\hline
$(-, -, -, -)$ &  $\Phi^a$ & $\sqrt {(2n+2)^2/R_1^2+ (2m+2)^2/R_2^2}$\\
\hline
$(-, +, -, +)$ &  $\Phi^{\hat{a}}$ & $\sqrt {(2n+1)^2/R_1^2+(2m+1)^2/R_2^2}$\\
\hline
$(P^y=+, P^{y'}=+)$ & $H$ & $2n/R_1$ \\
\hline
$(P^y=-, P^{y'}=-)$ & $H^c$ & $(2n+2)/R_1$ \\ 
\hline
\end{tabular}
\end{center}
\end{table}

\renewcommand{\arraystretch}{1.4}
\begin{table}[t]
\caption{For the supersymmetric model $SU(5)$ on 
$M^4 \times S^1/(Z_2\times Z^\prime_z)\times S^1/(Z_2\times Z^\prime_2)$,
 the gauge superfields, the number of
4-dimensional supersymmetry and gauge symmetry on the 3-brane which
is located at the fixed point $(y=0, z=0),$ $(y=0, z=\pi R_2/2),$ $(y=\pi R_1/2, z=0)$, and 
$(y=\pi R_1/2, z=\pi R_2/2)$, or on the 4-brane which is located at the fixed line
$y=0$, $z=0$, $y=\pi R_1/2$, $z=\pi R_2/2$. We also include the fermions, left handed
quark doublet $Q$, right handed up-type quark $U$ and down-type quark $D$ that are on the
3-brane at $(y=\pi R_1/2, z=\pi R_2/2)$, the lepton doublet $L$, right handed
lepton $E$ and neutrino $N$ that are on the 3-brane at $(y=0, z=\pi R_2/2)$.
\label{tab:SUV11}}
\vspace{0.4cm}
\begin{center}
\begin{tabular}{|c|c|c|c|}
\hline        
Brane Position & Fields & SUSY & Gauge Symmetry\\ 
\hline
$(0, 0) $ &  $V^A_{\mu}$ & $N=1$ & $SU(5)$ \\
\hline
$(0, \pi R_2/2)$ & $V^a_{\mu}$, $\Sigma_6^{\hat a}$, $L$, $E$, $N$
  & N=1 & $SU(3)\times SU(2)\times U(1)$ \\
\hline
$(\pi R_1/2, 0) $ & $V^a_{\mu}$, $\Sigma_5^{\hat a}$  & N=1 & $SU(3)\times SU(2)\times U(1)$ \\
\hline
$(\pi R_1/2, \pi R_2/2) $ & $V^a_{\mu}$, $\Phi^{\hat a}$, $Q$, $U$, $D$
  & N=1 & $SU(3)\times SU(2)\times U(1)$ \\
\hline
$y=0$ & $V^A_{\mu}$, $\Sigma_6^A$  & N=2 & $SU(5)$ \\
\hline
$z= 0 $ & $V^A_{\mu}$, $\Sigma_5^A$  & N=2 & $SU(5)$ \\
\hline
$y=\pi R_1/2 $ & $V^a_{\mu}$, $\Sigma_5^{\hat a}$, $\Sigma_6^{a} $, $\Phi^{\hat a}$
  & N=2 & $SU(3)\times SU(2) \times U(1)$ \\
\hline
$z=\pi R_2/2 $ & $V^a_{\mu}$, $\Sigma_5^{a}$, $\Sigma_6^{\hat a} $, $\Phi^{\hat a}$
  & N=2 & $SU(3)\times SU(2) \times U(1)$ \\
\hline
\end{tabular}
\end{center}
\end{table}

We consider the space-time $M^4 \times S^1/(Z_2\times Z^\prime_z)
\times S^1/(Z_2\times Z^\prime_2)$, where the orbifold  
$S^1/(Z_2\times Z^\prime_z)
\times S^1/(Z_2\times Z^\prime_2)$  is defined by $S^1\times S^1$ moduloing
the equivalent classes
\bea
P_y ~:~ y \sim -y, ~~ P^\prime_y ~:~ y^\prime \sim -y^\prime, \nnb \\
P_z ~:~ z \sim -z, ~~ P^\prime_z ~:~ z^\prime \sim -z^\prime,
\eea
where $y^\prime = y-\pi R_1/2$ and $z^\prime =z- \pi R_2/2$. The physical
space is in the region: $0 \le y \le \pi R_1/2$ and
$0 \le z \le \pi R_2/2$. The detail set-up and discussions can be found
in Ref.~\cite{LTJ2}.

From the action Eq. (\ref{6daction}), we obtain that under
$P_y$ and $P_{y'}$, the vector multiplets transform as:
\bea
\label{parityy1}
& V(x^\mu,-y,z) = P_y V(x^{\mu},y,z) (P_y)^{-1}, ~~ V(x^\mu,-y^\prime,z)
= P^\prime_y V(x^{\mu},y^\prime,z) (P^\prime_y)^{-1}, \\
\label{parityy2}
&  \Sigma_5(x^\mu,-y,z) = - P_y \Sigma_5(x^\mu, y, z) (P_y)^{-1}, ~~
\Sigma_5(x^\mu,-y^\prime,z) = 
- P^\prime_y \Sigma_5(x^\mu, y^\prime,z) (P^\prime_y)^{-1}, \\
\label{parityy3}
&  \Sigma_6(x^\mu,-y,z) = P_y \Sigma_6(x^\mu, y, z) (P_y)^{-1}, ~~
\Sigma_6(x^\mu,-y^\prime,z) = 
 P^\prime_y \Sigma_6(x^\mu, y^\prime,z) (P^\prime_y)^{-1}, \\
\label{parityy4}
&  \Phi(x^\mu,-y,z) = - P_y \Phi(x^\mu, y, z) (P_y)^{-1}, ~~
\Phi(x^\mu,-y^\prime,z) = 
- P^\prime_y \Phi(x^\mu, y^\prime,z) (P^\prime_y)^{-1}. 
\eea
 For $P_z$ and
$P^\prime_z$,  the vector multiplet transformations
 are similar to those under $P_y$ and $P_{y'}$, {\it, i. e},
we just make the following transformation on subscripts:
$y \leftrightarrow z$ and $5 \leftrightarrow 6$.

We choose the following representations for $(P_y,P^\prime_y,P_z,P^\prime_z)$:
\bea
&& P_y=(+1,+1,+1,+1,+1), ~~ P^\prime_y=(+1,+1,+1,-1,-1),\\
&& P_z=(+1,+1,+1,+1,+1), ~~ P^\prime_z=(+1,+1,+1,-1,-1).
\eea

The KK mode expansions for the bulk fields can be found in Ref.~\cite{LTJ2}.
The parity assignment and particle spectrum for the gauge fields
and the Higgs fields on the 4-brane at $z=\pi R_2/2$
are given in Table 3, and the gauge 
superfields, the number of 4-dimensional
supersymmetry and gauge group on the 3-brane or 4-brane are shown
 in Table 4. 

It is clear that at the energy scale $\mu < 1/R_1$ and $ 1/R_2$, we get
the 4-dimensional $N=1$ effective theory described by the zero mode of $V^a$.
Similar to the reasons pointed out in section 2, we
assume that $10/R_1 \sim 1/R_2$. From the parity
assignment in the Table 3, at the energy scale $1/R_1 < \mu < 1/R_2$
we get a 5-dimensional $N=1$ effective theory described by the zero modes of
$V^a$ and $\Sigma^a_5$ along the $z$ direction. 
In order to avoid the proton decay, we put the quark
 fields ($Q$, $U$ and $D$)
on the 3-brane at $(y=\pi R_1/2, z=\pi R_2/2)$,  and the lepton and neutrino
fields ($L$, $E$, $N$) on the 3-brane at $(y=0, z=\pi R_2/2)$. 
Let us discuss the anomaly cancellation. The $SU(2)$ is a safe
Lie algebra. Because we put all the quark fields on the 
3-brane at $(y=\pi R_1/2, z=\pi R_2/2)$, there are no $SU(3)$ anomaly.
So, the possible anomaly must involve at least one $U(1)$.
In addition, the anomaly localized on the 3-brane at $(y=\pi R_1/2, z=\pi R_2/2)$
and the anomaly localized on the 3-brane at $(y=0, z=\pi R_2/2)$
have the opposite sign and same magnitude, and there are two ways to cancel
the anomaly:

(1) Similar to the discussions on 5-dimensional orbifold~\cite{anomaly},
we introduce the following Chen-Simons term on the 4-brane
(covering space-time $M^4\times S^1\times (z=\pi R_2/2)$)
 at $z=\pi R_2/2$~\footnote{For simplicity, 
we discuss the pure $U(1)$ case and the discussions for
the other cases are similar.}
\bea
{\cal L}_{CS} &=& -{1\over 4} {\beta \over {128 \pi^2}} 
\int_{M^4\times S^1 \times S^1} d^4x dy dz ~\delta (z-\pi R_2/2)~\theta (y)~
\epsilon_{MNOPQ} A^M F^{NO} F^{PQ}~, 
\eea
where $\beta$ is the constant to be adjusted to cancel
the anomaly, $\theta (y)=+1$ for $y \in (0, \pi R_1/2) \cup (\pi R_1, 3 \pi R_1/2)$,
and $\theta (y)=-1$ for $y \in (\pi R_1/2, \pi R_1) \cup (3 \pi R_1/2, 2 \pi R_1)$.

(2) We introduce the following topological term on the covering space-time
$M^4\times S^1 \times S^1$~\cite{top}
\bea
{\cal L} &=& - {1\over 8} {\beta \over {128 \pi^3}} 
\int_{M^4\times S^1 \times S^1}
 d^4x dy dz ~\epsilon_{LMNOPQ} \partial^L \theta (y, z) A^M F^{NO} F^{PQ}, 
\eea
where $\theta(y, z)$ satisfies the equation
\bea
[\partial_y, \partial_z] \theta  &=& 2 \pi \left( \delta (y) -\delta (y-\pi R_1/2)
+\delta (y-3 \pi R_1/2 ) -\delta (y- 2\pi R_1) \right)
\nonumber\\
&&
\times \left(\delta (z-\pi R_2/2)+ \delta (z-3 \pi R_2/2 )\right) ~.
\eea

Since the 6-dimensional bulk theory is basically 4-dimensional
 $N=4$ supersymmetric theory,
 we can not put the Higgs fields in the whole bulk. On the
other hand, we can put the Higgs doublets on the boundary 4-brane
at $z=\pi R_2/2$, which preserves the 4-dimensional $N=2$ supersymmetry.
The 5-dimensional Higgs superfield can be described by two 4-dimensional
chiral Higgs superfields, $H$ and $H^c$. The orbifold can project out one
of them on the boundary 3-brane, and makes
the 4-dimensional effective theory to be a chiral $N=1$ 
supersymmetric theory. As noted
in the last section,
in order to have the gauge coupling unification,
 we should be careful on how to put the Higgs fields in
high dimension. One might expect that putting one pair of Higgs doublets
on the 4-brane at $z=\pi R_2/2$  might work. However,
 noticing that at each KK level
with mass $(2n+2)/R_1$, we have four chiral Higgs doublet fields,
 $V^a$ and $\Sigma^a_5$. Then, the beta functions are
$(b_1,b_2,b_3)=(-6/5,2,6)$ and $(b_2-b_1,b_3-b_2)=(16/5,4)$. Comparing 
with the MSSM beta function difference $(b_2-b_1,b_3-b_2)=(28/5,4)$, we find that
the above suggestion is impossible to unify the gauge couplings.
To rescue from this problem, we may put two extra singlets with hypercharge $1$
on the 4-brane at $z=\pi R_2/2$,
 and the beta functions become $(b_1,b_2,b_3)=(-18/5,2,6)$.
It just recovers what we have in the MSSM, {\it i.e.},
$(b_2-b_1,b_3-b_2)=(28/5,4)$, and can work as expected.
For example, assuming $1/R_2=20/R_1$ and $1/R_1 = 10$ TeV, we can achieve 
the gauge coupling unification at around $530$ TeV with the MSSM threshold
scale $M_{SUSY}$ varying $200-1000$ GeV. Just as the case in 4-dimension, the
uncertainty of the boundary conditions given by the supersymmetry threshold gives
only small corrections.

Another choice is that we put only one Higgs doublet on the 4-brane at $z=\pi R_2/2$,
which may be identified as 4-dimensional $H_d$ and its mirror partner $H^c_d$. 
The chiral superfield
$H_u$ can be put on the 3-brane at $(y=\pi R_1/2,z=\pi R_2/2)$.
The beta function is $(b_1,b_2,b_3)=(-3/5,3,6)$ and the difference 
among them is
$(b_2-b_1,b_3-b_2)=(18/5,3)$. As observed in~\cite{ddg}, this
choice can lead to the gauge coupling unification. Assuming
$1/R_2 = 20/R_1$ and $1/R_1 =10$ TeV, we obtain that
the unification scale $M_U$ is about $770$ TeV with $M_{SUSY}$ varying
from $200-1000$ GeV. Similar to above, the anomaly from the Higgs field $H_d$
and $H_d^c$ on the 4-brane can be cancelled by introducing the suitable
Chen-Simons terms on the 4-brane at $z=\pi R_2/2$
or the topological term in the bulk. However, the anomaly from the 
chiral superfield $H_u$ can not
be cancelled unless we introduce another Higgs doublet, say ${\tilde H}^c_u$,
on the 3-brane at $(y=\pi R_1/2,z=\pi R_2/2)$ or $(y=0,z=\pi R_2/2)$.
Of course, we have to make ${\tilde H}^c_u$  massive and let it disappear in the
effective 4-dimensional theory below the compactification scale.

On the 3-brane at $(y=\pi R_1/2,z=\pi R_2/2)$, the
 field $\Phi^{\hat a}$ $(3,2,-\frac{5}{6})$ does not
vanish. So, we have the localized superpotential $H_u D^c \Phi^{\hat a}$
on the 3-brane at $(y=\pi R_1/2,z=\pi R_2/2)$.
Since the $U(1)$ charge has to be balanced on the 3-brane localized Yukawa
superpotential and we must have the anomaly cancellation,
 we then have the charge quantization from the consistent conditions
of these models.

\section{Discussions and Conclusion} \label{sec4}

\begin{figure}[t]
\centerline{\psfig{figure=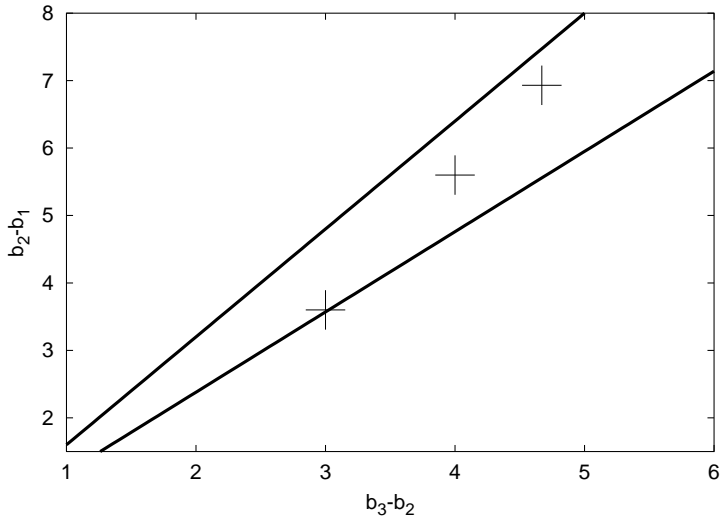,height=7cm,width=10cm}}
\caption{\small $b_3-b_2$ versus $b_2-b_1$. The allowed region by
gauge coupling unification for the supersymmetric models
is in between the two lines with slopes
$1.19$ and $1.60$. Similar results hold for the non-supersymmetric model.
 }
\label{fig}
\end{figure}

First, as noted before, we concentrate on the scenarios with 
$1/R_2 \sim 10/R_1$
in order to suppress the possible uncertainties on the gauge coupling
unification. For the case with $1/R_2 \sim 1/R_1$, the dominating radiative
corrections given by the massive KK excitations are basically $SU(5)$
symmetric and the relative runnings are given by the subleading
contributions which might suffer the uncertainties from the
loop corrections to the masses of the KK states, the supersymmetry
breaking soft parameters and the Higgs mechanism.
However, these uncertainties deserve further and careful study.
Because if these uncertainties are indeed small, we can
construct many new models with low energy gauge unfication,
for example, the 6-dimensional $N=1$ or $N=2$
 supersymmetric $SU(5)$ model on the 
space-time $M^4\times T^2/Z_6$, and the
6-dimensional $N=1$ or $N=2$ supersymmetric $SU(5)$, $SU(6)$,
$SO(10)$ and $E_6$ models on the 
space-time $M^4\times D^2$ or $M^4\times A^2$ where $D^2$ is the disc and
$A^2$ is the annulus.

Second, we learned from the discussions in sections \ref{sec1} and 
\ref{sec2} that
basically we only need check $(b_2-b_1,b_3-b_2)$ of the 5-dimensional KK
excitations to examine whether the particle content can lead to
the gauge coupling unification. In particular, as shown in the section \ref{sec2}
checking in this way is also reliable even in the 6-dimensional
setup if $1/R_2 \sim 10/R_1$ involved.
Because of the orbifold projection,  parts of the
$SU(5)$ representations disappear in the KK states of masses $(n+1)/R_1$
or $2(n+1)/R_1$ and this makes the power one contributions
survive from leading power two $SU(5)$ symmetric contributions 
(actually the power two contributions vanish 
in the supersymmetric model because of the $N=4$ supersymmetry).
If $1/R_2 \sim 10/R_1$ (or $\Lambda R_1 \sim 10 \Lambda R_2$) involved,
the $SU(5)$ asymmetric power one contribution is then at the leading
order which can suppress the possible uncertainties on the gauge
coupling unification. 
In this framework, we have several scenarios that do have low energy
unification: (I) the non-supersymmetric $SU(5)$ scenario
with three pairs of bulk Higgs doublets where the differences among
the beta functions are $(b_2-b_1,b_3-b_2)=(104/15,14/3)$; 
(II) the supersymmetric $SU(5)$ scenario with one
pair of Higgs doublets and one pair of hypercharge one singlets
on the boundary 4-brane at $z=\pi R_2/2$, where the differences among
the beta functions are $(b_2-b_1,b_3-b_2)=(28/5,4)$;
(III) the supersymmetric $SU(5)$ scenario with one Higgs doublet
on the boundary 4-brane at $z=\pi R_2/2$, where the the differences among
the beta functions are $(b_2-b_1,b_3-b_2)=(18/5,3)$.
We remind the readers that $b_2-b_1$ and $b_3-b_2$ satisfy
a roughly linear dependence, {\it i.e.}, the larger $b_3-b_2$, the
larger $b_2-b_1$ is needed. By the way, we would like to emphasize that
 two natural scenarios do not have the gauge unification, and they are:
(1) for the non-supersymmetric case, $SU(5)$ gauge fields and one pair of
Higgs doublets (or one Higgs doublet) in the bulk;
(2) for the supersymmetric case, $SU(5)$ gauge fields in the bulk and
one pair of Higgs doublets on the boundary 4-brane at $z=\pi R_2/2$.

This phenomenon is not strange to us in view of the fact
that the dominating one-loop running equation is actually
linearly dependent on the beta functions, as can be seen
in Eq.s (\ref{run1}), (\ref{run2}), (\ref{run23}) 
 and (\ref{run3}). To achieve the gauge coupling
unification, we should adjust the particle content so that the
more rapid the relative running between $\alpha^{-1}_3$
and $\alpha^{-1}_2$, the more rapid the relative running between
$\alpha^{-1}_2$ and $\alpha^{-1}_1$, as we need to obtain
the gauge unification. If one of them becomes
slower, so does the other to get the gauge
coupling unification. In the Figure \ref{fig}, we give
a plot on the correlated region for $b_2-b_1$ and $b_3-b_2$.
We may understand the slope by realizing that
\bea
\frac{\alpha^{-1}_1(m_Z)-\alpha^{-1}_2(m_Z)}
{\alpha^{-1}_2(m_Z)-\alpha^{-1}_3(m_Z)} \approx 1.38.
\eea
Using Eq. (\ref{run1}) or (\ref{run3}), we can translate this number
to be the required ratio $r=(b_2-b_1)/(b_3-b_2)$. Taking the
values of $\alpha_i$'s at $10$ TeV instead and assuming the
$M_{SUSY}$ varies in the range $200-1000$ GeV, we can get the required
ratio, $r$. Further assuming that the mismatch at the $M_U$ scale,
 $\delta=(\alpha^{-1}_3-\alpha^{-1}_2)/\alpha^{-1}_2$, is less than
$5\%$, and the unified $\alpha^{-1}$ is smaller than $50$, we obtain the 
region which is allowed by the gauge
 unification: $1.19 < r < 1.60$. (Without supersymmetry,
 the bound is roughly
$1.15 <r <1.52$ at $10$ TeV scale and $1.20 < r < 1.56$ at $1$ TeV scale). 
The two lines in the Fig. \ref{fig} correspond to these two bounds. 
The above successful scenarios have the following ratios:
(I) $r=(104/15)/(14/3)\approx 1.49$; (II) $r=28/5/4=1.4$;
(II) $r=(18/5)/3=1.2$. And these three $r$ are plotted as
three points in the Fig. \ref{fig}.

One intersting question still remains, which is whether we can 
find a natural and simple  scenario in high dimension which 
can unify the gauge couplings without adding 
the exotic particles or extra Higgs by hand.
According to our experiences, this is a really hard question.

Furthermore, there exists another kind of scenarios that might deserve 
further study. We can use the Higgs mechanism to break the GUT gauge
symmetry and assume that the
GUT breaking scale is much larger than the scale of $1/R_1$ and $1/R_2$.
And we can use the orbifold projection to forbid the proton decay operators,
and use the KK states to accelerate the gauge unification. Of course,
 the uncertainties on the gauge unification are avoided.
The non-supersymmetric models works similarly, and we might discuss
any GUT models on the space-time $M^4\times [S^1/(Z_2\times Z_2')]^n$, or
 $M^4\times D^2$,  or $M^4\times A^2$, for example $SO(10)$ and $E_6$.
However, it seems to us that we might not construct the supersymmetric extensions
of this kind of models, the key points are: (1) 
the Higgs, which break the gauge symmetry, can not break the parity,
{\it i. e.}, the neutral Higgs which has VEV must transform
trivially under the discrete symmetry which acts on the extra
space manifold; (2) the Higgs on the boundary brane can not
give the large masses (compare to $1/R$) to the bulk fields by Higgs
mechanism, although they might change the boundary conditions for the bulk 
fields~\cite{CLP};
(3) the 4-dimensional $N=2$ hypermultiplets only have gauge interactions~\cite{MFS}.

In short, because of the problems arising from the fermion unification in
the traditional Grand Unified Theory and the mass hierarchy
between the 4-dimensional Planck scale and weak scale, 
we suggest the low energy gauge unification theory
with low high-dimensional Planck scale. We discuss
 the non-supersymmetric $SU(5)$ model on 
the space-time $M^4\times S^1/Z_2 \times S^1/Z_2$
where there are three pairs of Higgs doublets in the bulk;
the supersymmetric $SU(5)$ model on 
the space-time $M^4\times S^1/(Z_2\times Z_2')
\times S^1/(Z_2\times Z_2')$ where there are one
pair of Higgs doublets and one pair of hypercharge one singlets
on the boundary 4-brane at $z=\pi R_2/2$, or one Higgs doublet ($H_d$)
on the boundary 4-brane at $z=\pi R_2/2$. The $SU(5)$ gauge symmetry is broken
by the orbifold projection for the zero modes, and the gauge unification
is accelerated due to the $SU(5)$ asymmetric light KK states.
In our models, we forbid the proton decay, still keep the
charge quantization, and automatically solve the fermion mass
problem. We also comment on the anomaly cancellation and
the other possible scenarios for
low energy gauge unification.

\section{ Acknowledgments} 
This work was supported in part by
the U.S.~Department of Energy under Grant No.~DOE-EY-76-02-3071.

\end{document}